\documentclass[twocolumn,showpacs,preprintnumbers,amsmath,amssymb]{revtex4}
\usepackage{graphicx}
\usepackage{dcolumn}
\usepackage{bm}

\newcommand{\beq}{\begin{equation}}
\newcommand{\eeq}{\end{equation}}
\newcommand{\beqa}{\begin{eqnarray}}
\newcommand{\eeqa}{\end{eqnarray}}
\newcommand{\mat}[1]{{\rm \mathbf{#1}}}
\newcommand{\op}[1]{{\hat #1}}
\newcommand{\ve}[1]{{\bm #1}}

\newcommand{\dd}{{\rm d}}

\begin{document}

\title{Faster than Lyapunov decays of classical Loschmidt echo}

\author{Gregor Veble$^{1,2}$ and Toma\v z Prosen$^{1}$}
\affiliation{$^{1}$Physics Department, FMF, University of Ljubljana, Ljubljana, Slovenia\\
$^2$ Center for Applied Mathematics and Theoretical Physics, University of Maribor, Maribor, Slovenia}

\date{\today}

\begin{abstract}
We show that in the {\em classical interaction picture} the echo-dynamics, namely the composition 
of perturbed forward and unperturbed backward hamiltonian evolution, 
can be treated as a time-dependent hamiltonian system. For strongly chaotic (Anosov) systems
we derive a cascade of exponential decays for the classical Loschmidt echo,
starting with the leading Lyapunov exponent, followed by a sum of two largest exponents, 
etc. In the loxodromic case a decay starts with the rate given as twice the largest Lyapunov exponent.
For a class of perturbations of symplectic maps the echo-dynamics       
exhibits a drift resulting in a super-exponential decay of the Loschmidt echo.
\end{abstract}

\pacs{05.45.Ac,05.45.Mt,03.67.Lx}

\maketitle

Analyzing the parametric stability of quantum dynamics through the
so-called {\em fidelity} or {\em Quantum Loschmidt Echo} (QLE) has become increasingly 
popular and useful tool, either in the context of experiments with many-body 
systems \cite{Usaj}, quantum computation \cite{Qcomp}, or
simple dynamical systems \cite{Peres,Jalabert,Prosen,PZ,Tomsovic,Beenakker}.
The corresponding {\em Classical Loschmidt Echo} (CLE) has been defined 
\cite{PZ,BC1,Eckhardt,Veble}, as
\begin{equation}
F(t) = \int \dd^N\ve{x} \rho(\ve{x})\rho_{\rm E}(\ve{x},t)
\label{eq:CLE}
\end{equation}
where $\rho(\ve{x})$ is an $L^2$ normalized non-negative initial density in 
$N=2d$-dimensional classical phase space with coordinates $\ve{x}$,
and $\rho_{\rm E}(\ve{x},t)$ is its image after the {\em echo-dynamics},
i.e. a composition of a hamiltonian flow with a slightly perturbed time-reversed
hamiltonian flow. It has been found that for classically 
chaotic systems CLE follows QLE only for a short time that scales as $\log\hbar$.
For longer times, CLE of chaotic systems has been shown to follow time
correlation functions \cite{Veble}.
However, CLE is in itself an interesting quantity in classical statistical mechanics 
as it provides a way to quantify the old Loschmidt-Boltzmann controversy.

For shorter times, before the relaxation of the initially localized density
under echo-dynamics takes place, it has been found numerically\cite{BC1} that CLE decays
exponentially with the rate given by the Lyapunov exponent. No classical
mechanism for this phenomenon has been given, apart from the necessary
correspondence with QLE for which a semiclassical theory of Lyapunov decay
exists \cite{Jalabert}. 
In this letter we report several surprising analytical 
results for the case of localized initial densities and for sufficiently weak perturbations: 
(i) In many-dimensional systems, a cascade of Lyapunov decays is predicted, with the 
exponents which are given as consecutive sums of largest few Lyapunov exponents.
Hence precise conditions for previously observed simple Lyapunov decay are understood.
(ii) In loxodromic case of degenerate largest Lyapunov exponent 
$\lambda_1=\lambda_2$ we find initial decay of CLE with exponent $2\lambda_1$. 
(iii) For maps under special conditions 
we find non-zero average drift of the echo-dynamics
resulting in faster than exponential decay of CLE.
The same results should apply for quantum fidelity (QLE) up to the
log-time, namely until the Wigner function follows the
classical density.

The propagation of classical densities in phase space is governed
by the unitary Liouville evolution $\op{U_\delta(t)}$
\beq
\frac{\dd}{\dd t} \op U_\delta(t)={\cal \op L}_{H_\delta(\ve{x},t)} \op U_\delta(t)
\label{eq:dUdt}
\eeq
where
${\cal \op  L}_{\! A(\ve{x},t)} =  
\left(\ve \nabla\! A(\ve{x},t)\right)\cdot\mat J \ve \nabla,$ 
$A$ is any observable, and
\beq
H_\delta(\ve{x},t) = H_0 (\ve{x},t) + \delta V(\ve{x},t),
\label{eq:ham}
\eeq
is a generally time-dependent family of Hamiltonians with perturbation
parameter $\delta$. Matrix $\mat J$ is the usual symplectic unit.
Similarly,
$
\dd\op U^\dagger_\delta(t)/\dd t= - \op U^\dagger_\delta (t) 
{\cal \op  L}_{H_\delta(\ve{x},t)}.
$
The {\em classical echo propagator}
that composes perturbed forward evolution with the unperturbed backward
evolution is also unitary and is given by
\beq
\op U_{\rm E}(t)=\op U^\dagger_0(t)\,\op U_\delta (t) \label{eq:Md}.
\eeq
Using eqs. (\ref{eq:dUdt},\ref{eq:ham},\ref{eq:Md}) and writing 
$\op{U}_\delta(t) = \op{U}_0(t) \op{U}_{\rm E}(t)$ we get
\beq
\frac{\dd}{\dd t} \op U_{\rm E} (t)= \left\{ \op U_0^\dagger(t)
{\cal \op L}_{\delta V(\ve{x},t)}\op U_0(t)\right\} \op U_{\rm E} (t). 
\label{eq:timeder}
\eeq
The classical dynamics has the nice property that the evolution
is governed by characteristics that are simply the classical phase
space trajectories, so the action of the evolution operator on any
phase space density is given as
$
\op U_0(t)~\rho=\rho \circ \ve \phi_t^{-1},
$
where $\phi_t^{-1}$ denotes the backward (unperturbed) phase space flow
 from time $t$ to $0$. Similarly, 
$
\op U_0^\dagger(t)~\rho=\rho \circ \ve \phi_t,
$
where $\phi_t$ represents the forward phase space flow
from time $0$ to time $t$. Here and in the following we assume the 
dynamics to start at time $0$.

We note that echo-dynamics (\ref{eq:Md}) can be treated as
Liouvillian dynamics in {\em interaction picture}, since
\beqa
&&\left\{ \op U_0^\dagger(t) {\cal \op L}_{A(\ve{x},t)} \op U_0 (t) \rho \right\}(\ve x)= \\&&
=\op U_0^\dagger(t)\left(\ve \nabla_{\ve{x}}  A(\ve x,t)\right) \cdot \mat J \ve \nabla_{\ve x}
 \rho\left(\ve \phi_t^{-1}(\ve x)\right)=\nonumber\\
&&
=
\left(\ve\nabla_{\ve{\phi}_t(\ve{x})} A (\ve \phi_t(\ve x),t)\right)
\cdot \mat J \ve \nabla_{\ve{\phi}_t(\ve{x})} \rho\left(\ve \phi_t^{-1}\left(\ve \phi_t(\ve x)\right)\right)
=\nonumber\\
&&
=
\left(\ve\nabla_{\ve{x}} A (\ve \phi_t(\ve x),t)\right)
\cdot \mat J \ve \nabla_{\ve{x}} \rho(\ve x)
=\left\{ {\cal \op L}_{A\left(\ve \phi_t(\ve x),t\right)} ~\rho\right\}({\ve x}). \nonumber
\eeqa
In the last line we used the invariance of the Poisson bracket under the flow. 
This extends eq.~(\ref{eq:timeder}) to form (\ref{eq:dUdt})
\beq
\frac{\dd}{\dd t} \op U_{\rm E} (t)= 
{\cal \op L}_{H_{\rm E}(\ve x,t)}~\op U_{\rm E} (t) 
\eeq
where the {\em echo Hamiltonian} is given by 
\beq
H_{\rm E}({\ve x},t) = \delta V\left(\ve \phi_t(\ve x),t \right). \label{eq:Hf}
\eeq
The function $H_{\rm E}$ is nothing but the perturbation
part $\delta V$ of the original Hamiltonian, which, however, is evaluated
 at the point that is 
obtained by forward propagation with the unperturbed original Hamiltonian. 
It is important to stress that this 
is not a perturbative result but an exact expression. Also, even if the 
original Hamiltionian system was time independent, the echo dynamics
obtains an explicitly time dependent form.
Trajectories of the echo-flow are given by Hamilton equations
\beq
\dot {\ve x}= \mat J~ \ve \nabla H_{\rm E} (\ve x,t). 
\label{eq:hameq}
\eeq
At this point we limit our discussion only to time independent original 
Hamiltonians and perturbations.
Slightly more general case of periodically driven systems reducible to
symplectic maps shall also be discussed later.

Inserting (\ref{eq:Hf}) into eq. (\ref{eq:hameq}) yields
\beq
\dot {\ve x} = \delta \mat J \ve \nabla_{\ve x} V(\ve \phi_t(\ve x)) =
\delta \mat J {\mat M_t^T(\ve x)} (\ve \nabla V)(\ve \phi_t(\ve x)) \label{eq:dyn}.
\eeq
Here we have introduced the {\em stability matrix} $\mat M_t (\ve x)$,
$[\mat M_t(\ve x)]_{i,j} = \partial_j [\ve\phi_t(\ve x)]_i$.
From now on we assume that the flow $\ve{\phi}_t$ is Anosov.
To understand the dynamics (\ref{eq:dyn}) we need to explore the properties of 
$\mat M_t$. We start by writing the matrix
$
\mat M_t^T(\ve x) \mat M_t(\ve x)=
\sum_j e^{2 \lambda_j t} d_j^2(\ve x,t)
 \ve v_j(\ve x,t)\otimes \ve v_j(\ve x,t)
$
expressed in terms of orthonormal 
eigenvectors $\ve v_j(\ve x,t)$ and eigenvalues
$d_j^2(\ve x,t) \exp(2 \lambda_j t)$. After the ergodic time $t_{\rm e}$
necessary for the echo trajectory to explore the available region of phase space, 
Osledec theorem\cite{Osledec} guarantees 
that the eigenvectors of this matrix converge to Lyapunov eigenvectors being
independent of time, while $d_j(\ve x,t)$ grow slower than exponentially, 
so the leading exponential growth defines the Lyapunov exponents $\lambda_j$.
Similarly, the matrix
$
\mat M_t(\ve x) \mat M_t^T(\ve x) =\sum_j e^{2 \lambda_j t} c_j^2(\ve x_t,t)
 \ve u_j(\ve x_t,t)\otimes \ve u_j(\ve x_t,t), \nonumber
$
where $\ve x_t=\ve\phi_t(\ve x)$, has the same eigenvalues 
[$c^2_j(\ve x_t,t) \equiv d^2_j(\ve x,t)$], and its eigenvectors depend on the 
final point $\ve x_t$ only,
as the matrix in question can be related to the backward evolution. 
The vectors $\{\ve u_j(\ve x_t)\}$, $\{\ve v_j(\ve{x})\}$, 
constitute left, right, part, respectively, of the {\em singular value decomposition} of 
$\mat M_t(\ve{x})$, so we write for $t \gg t_{\rm e}$
\beq
\mat M_t(\ve x)=
 \sum_{j=1}^N \exp(\lambda_j t) 
~\ve e_j(\ve \phi_t(\ve x))
\otimes  \ve f_j(\ve x) 
\label{eq:Mt}
\eeq
assuming that the limits 
$\ve e_j(\ve x) = \lim_{t\to\infty} c_j(\ve x,t)\ve u_j (\ve x,t)$,
$\ve f_j(\ve x) = \lim_{t\to\infty} d _j(\ve x,t) \ve v_j(\ve x,t)$
exist. Rewriting eq. (\ref{eq:dyn}) by means of eq. (\ref{eq:Mt})
we obtain
\beq
\dot {\ve x}=
\delta \sum_{j=1}^N \exp(\lambda_j t) 
~W_j(\ve \phi_t(\ve x))~
 \ve h_j(\ve x) 
\label{eq:dyn1}
\eeq
where $\ve h_j(\ve x)=\mat J\ve f_j(\ve x)$, and introducing new observables
\beq
W_j(\ve x)=
\ve e_j(\ve x)\cdot \ve \nabla V(\ve x).
\eeq

At this point it is perhaps necessary to discuss the nature of the
vector fields $\ve e_j(\ve x)$, $\ve f_j(\ve x)$
\cite{remark}. While the theorem \cite{Osledec}
guarantees the existence of these directions, the actual
sizes of these fields as given by $c_j$, $d_j$ are 
not yet well understood. Numerical data suggest that
these quantities do converge as $t\to\infty$ for an Anosov system.

For small perturbations the echo trajectories
remain close to initial point $\ve x(0)$ for times large in comparison to the
internal dynamics of the system ($t_{\rm e}$, Lyapunov times, decay of correlations, etc), 
and in this regime
the echo evolution can be linearly decomposed along different independent
directions $\ve h_j(\ve x(0))$ 
\beq
\ve x(t)=\ve x(0) +\sum_{j=1}^N y_j(t)
 ~\ve h_j(\ve x(0)).
\label{eq:decomp}
\eeq
For longer times, the point $\ve x(t)$ moves away from the initial point,
but the dynamics is still governed by the local unstable vectors at the 
evolved point. Therefore the decay of fidelity is governed by the
spreading of the densities along the conjugated unstable manifolds 
defined by the vector field $\ve h(\ve x)$.

Inserting (\ref{eq:decomp}) into (\ref{eq:dyn1}) we obtain for each direction
$\ve h_j$
\beq
\dot y_j =  \delta ~ \exp(\lambda_j t) W_j(\ve \phi_t(\ve x)).
\label{eq:yyy}
\eeq
For {\em stable} directions with $\lambda_j<0$, clearly after a certain
time the variable $y_j$ becomes a constant of the order $\delta$.

For {\em unstable} directions with $\lambda_j>0$, we
introduce $z_j$ as $y_j = \delta \exp(\lambda_j t) z_j$ and rewrite the above
equation as
\beq
\dot z_j + \lambda_j z_j = W_j(\ve \phi_t(\ve x)).
\label{eq:damp}
\eeq
The right hand side of this equation is simply the evolution of the
observable $W_j$ starting from a point in phase space $\ve{x}=\ve{x}(0)$. 
Due to assumed ergodicity of the flow $\ve{\phi}_t$, $W_j(\ve \phi_t(\ve x))$
has well defined and {\em stationary} statistical properties such as
averages and correlation functions. Thus the solution $z_j(t)$ of the
linear damped equation (\ref{eq:damp}) has also stationary statistics and
well defined time- and $\delta$-independent 
probability distribution $P_j(z_j)$. Its moments can be expressed 
in terms of moments and correlation functions
of $W_j$, in particular $\overline{W_j}=0$ \cite{firstmoment}.
The analysis remains valid in a general case of explicitly time-dependent $W_j$
\cite{remark}.

Going back to the original
coordinate $y_j$ we obtain its distribution as
$K_j(y_j)=P_j(z_j)\dd z_j/\dd y_j$,
or
$
K_j(y_j)=P_j\left(\exp(-\lambda_j t) y_j/\delta\right)
\exp(-\lambda_j t)/\delta.
$
This probability distribution tells us how, on average, points
within some initial (small) phase space set 
of characteristic diameter $\nu$ spread 
along locally well defined unstable Lyapunov direction $j$
and  therefore represents an averaged
 kernel of the evolution of such densities along
this direction. Starting from the initial 
localized density $\rho_0$, of small width $\nu$
such that the decomposition (\ref{eq:decomp}) does
not change appreciably along $\rho_0$, the echo dynamics for 
densities solves as
$
\rho_t(\ve y)=\int \dd^N \ve{y}'
\rho_0(\ve y^\prime)
\prod_j K_j(y_j-y_j^\prime).
$
For stable directions $j$ we set $K_j(y_j) = \delta(y_j)$, as the 
shift of $y_j$ (of order $\delta$) can be neglected as compared to 
unstable directions. This also implies that the assumption
$\delta \ll \nu$ is necessary in order to get any echo at
all after not too short times. CLE (\ref{eq:CLE}) can now be written
as
$F(t)=\int \dd^N\ve y ~\rho_0(\ve y) ~\rho_t(\ve y)$.
As long as the width $\nu_j$ of $\rho_0$ along the unstable 
direction $j$ is much
larger than the width of the kernel $K_j$, there is no appreciable
contribution to the fidelity decay in that direction. At time
\beq
t_j=(1/\lambda_j)\log(\nu_j/(\delta\gamma_j)),
\eeq 
where $\gamma_j$ is a typical width of the distribution $P_j$, the width of
the kernel is of the order of the width of the distribution along the
chosen direction. After that time, the overlap between the two
distributions along the chosen direction
starts to decay with the same rate as the value of
the kernel in the neighborhood of $y_j=0$, which is
$\propto \exp(-\lambda_j t)$. The total overlap decays as
\beq
F(t)\approx \prod_{j;~t_j<t} \exp\left[-\lambda_j (t-t_j)\right], 
\label{eq:multilyap}
\eeq
where only those unstable directions contribute to the decay for
which $t_j>t$. As the time $t_j$ is shorter the higher the corresponding
Lyapunov exponent $\lambda_j$, fidelity will initially decay with
the largest Lyapunov exponent $\lambda_1$. In chaotic 
systems with more than two degrees of freedom we, however,
 expect to observe an increase of decay rate after the time
$t_2$, etc.
Eq. (\ref{eq:multilyap}) provides
good description for CLE as long as $F(t)$ does not approach the
saturation value $F_\infty \sim \nu^N$ where the asymptotic decay
of CLE is then given by leading Perron-Frobenius eigenvalue \cite{Veble}.

\begin{figure}
\centerline
{
\includegraphics[width=4.2cm]{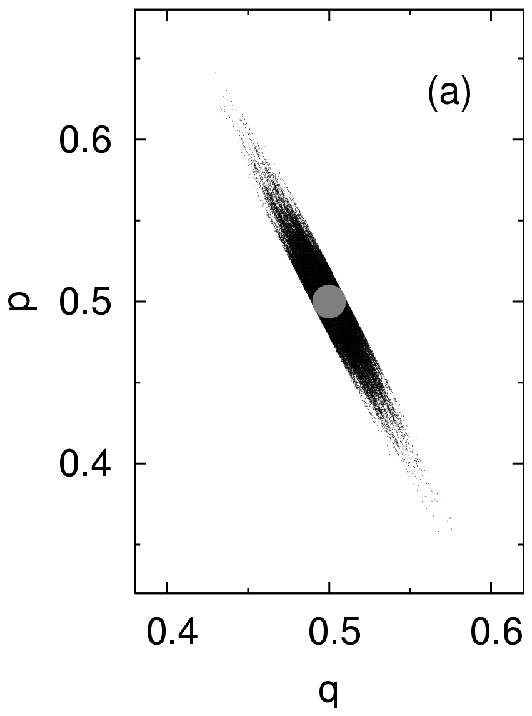}\includegraphics[width=4.2cm]{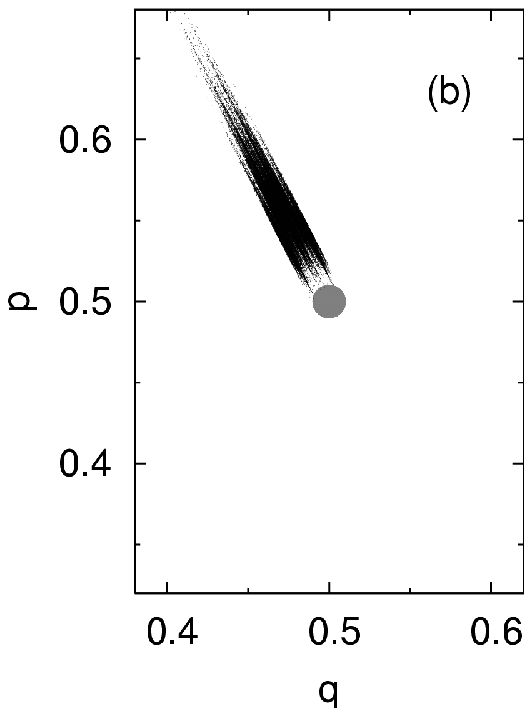}
}
\caption{Density evolution of echo-dynamics 
for the perturbed cat map [eq.~(\ref{eq:cat}), $\mu=0.3$]. The initial density is a
characteristic function on the circle (gray) centered at $(q,p)=(0.5,0.5)$
having the radius $0.01$, while the evolved density at the time 
$t=21$ is represented by $10^5$ dots. 
Figures (a,b) refer to corresponding cases for perturbations without and with
drift, respectively ($\delta=10^{-9}$, see text for details).
\label{fig:1}}
\end{figure}

\begin{figure}
\centerline{\includegraphics[width=7.8cm]{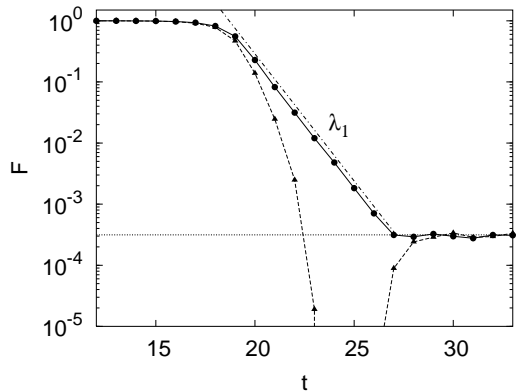}}
\caption{CLE as a function of time for the same conditions as in fig.~\ref{fig:1} 
except using $10^6$ points. The circles refer to the case (a) (no drift), chain 
line is the theoretical Lyapunov decay with $\lambda_1=0.958$, and the triangles refer to the ballistic 
case (b) (drift). In both cases the fidelity
saturates at the plateau (dotted line) given by the relative volume (area) of the
initial set.
\label{fig:2}}
\end{figure}

\begin{figure}
\centerline{\includegraphics[width=7.8cm]{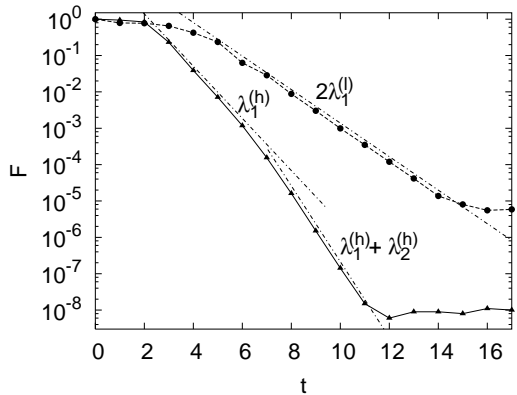}}
\caption{CLE for two examples of 4D cat maps perturbed as explained in text.
Triangles refer to doubly-hyperbolic case where initial set was a 4-cube
$[0.1,0.11]^4$, and $\delta=2\cdot 10^{-4}$, whereas circles
refer to loxodromic case where initial set was $[0.1,0.15]^4$, and 
$\delta=3\cdot 10^{-3}$. In both cases initial density was sampled by $10^9$ points.
Chain lines give exponential decays with theoretical rates, 
$\lambda_1=1.65,\lambda_1+\lambda_2=2.40$ (doubly-hyperbolic),
and $2\lambda_1=1.06$ (loxodromic).
\label{fig:3}}
\end{figure}

Though the above theory has been developed for smooth flows, the generalization
to ergodic symplectic maps on bounded phase space ${\cal M}$ is 
straightforward. We adopt notation of Ref.~\cite{PZ}, sect.~4:
$\ve\phi \equiv \ve\phi_1$, discrete time $t$ is an integer and a general small 
perturbation is given by a composition 
$\ve{\phi}_\delta = \ve{\phi}\circ\ve{g}_\delta$ with a
near-identity map $\ve{g}_\delta$ generated by a vector field $\ve{a}(\ve{x})$,
$\dd \ve{g}_\delta/\dd\delta = \ve{a}(\ve{g}_\delta)$ with initial condition
$\ve{g}_0(\ve{x})=\ve{x}$. For the perturbed map to remain symplectic we write 
$\ve a(\ve x)=\mat{J}\ve{\nabla} V(\ve{x})$ for some potential $V(\ve{x})$.
We note that on compact phase space $V(\ve{x})$ does not need to be unique and 
continuous, e.g. for a unit 2-torus $V(q+1,p) = V(q,p)+\alpha,V(q,p+1)=
V(q,p)+\beta$ where $\alpha,\beta$ are arbitrary constants. Provided hyperbolic
orbits with inversion do not exist, $\eta\equiv 0$ \cite{remark},
one finds a non-vanishing drift of echo-dynamcis, $\overline{W}_j\neq 0$, 
resulting in a possible super-exponential decay of CLE 
if the mean $\overline{W_j}/\lambda_j$ of the distribution $P_j(z_j)$ is larger 
than its width $\gamma_j$.

In order to illustrate super-exponential versus exponential decay of CLE 
we consider the perturbed cat map
\beqa
\overline p&=&p+q-\frac{\mu}{2 \pi} \sin(2\pi q)+\delta s(q), \quad \mu\in[0,1)\nonumber\\
\overline q&=&q+\overline p. \label{eq:cat}
\eeqa
The perturbation was chosen either as: (a)
$s(q)= \sin(2\pi q)/(2\pi)$, or (b) $s(q)=1/(2 \pi)$ where
the perturbation is a shift in phase space. The main difference between
the two is that the case (a) corresponds to zero drift since a unique
smooth potential exists ($\alpha=\beta=0$), 
while in case (b) $\ve{a}={\rm const}\neq 0$ and 
the Lyapunov fields 
have  a predominant direction in phase space for this system,
e.g. for the unperturbed cat map ($\mu=0$) $\ve e(\ve {x}), \ve f(\ve{x}), \ve h(\ve{x})$ are constant.
Since the perturbed cat has no orbits with inversion\cite{remark} the corresponding
phase space observable $W_j$ for the case (b) has a distinct nonzero
average value, causing the kernel $K_j$ to drift exponentially in time.
The difference in the qualitative nature of the two decays is shown in figure 
\ref{fig:1}. In figure \ref{fig:2} we show the behaviour of fidelity
as a function of time for the two cases, where a super-exponential
decay is observed in the case of drift.

Another result, which applies only to systems with two or more unstable
directions, is the occurrence of decays which are exponential but faster than
Lyapunov. In the case of well separated individual 
Lyapunov exponents the decay 
is expected to go through a cascade of increasing decay rates given by
(\ref{eq:multilyap}), whereas in the loxodromic case $\lambda_1=\lambda_2$ the rate is $2\lambda_1$.
We illustrate this numerically for 4D cat maps \cite{Rivas}:
$\ve{x}'= \mat{C} \ve{x} \pmod{1},$ $\ve x \in [0,1)^4$,
and
$$
\mat C_{\rm d-h}=\left[
\begin{array}{rrrr}
2 &-2 &-1 & 0 \\
-2& 3 & 1 & 0 \\
-1& 2 & 2 & 1 \\
2 &-2 & 0 & 1
\end{array}\right],\
\mat C_{\rm lox}=\left[
\begin{array}{rrrr}
0 & 1 & 0 & 0 \\
0 & 1 & 1 & 0 \\
1 &-1 & 1 & 1 \\
-1&-1 &-2 & 0
\end{array}\right]
$$
are two examples representing the doubly-hyperbolic and loxodromic case. 
Matrix $\mat C_{\rm d-h}$ 
has the unstable eigenvalues $\approx 5.22,2.11$,
while the large eigenvalues of $\mat C_{\rm lox}$ 
are $\approx 1.70 \exp(\pm {\rm i} 1.12)$. The perturbation for
both cases was done by performing an additional mapping at each timestep
$\bar{x}_1=x'_1 + \delta\sin\left(2\pi x_3\right) \pmod{1}$,
$\bar{x}_{2,3,4}=x'_{2,3,4}$.
In figure \ref{fig:3} we show the two types of decay which agree with
theoretical predictions.

In conclusion, we have developed a theory for short-time decay of CLE based on classical 
interaction picture. Our theory predicts several new phenomena,
in particular a cascade of exponential decays in systems with more than one unstable
direction and doubly-Lyapunov decay for the particular case of loxodromic stability.
Besides being related to quantum computation,
our results for systems with many degrees of freedom provide a way
to understand macroscopic irreversibility in classical statistical mechanics.

We acknowledge useful discussion with T.H.Seligman and 
financial support by the Ministry of Education, Science and Sport
of Slovenia, and in part by the U.S. ARO grant DAAD19-02-1-0086.


\begin{thebibliography}{99}
\bibitem{remark} We note that {\em signs} of the fields 
$\pm \ve{e}_j,\pm \ve{f}_j$ are not {\em unique} provided the flow has 
hyperbolic periodic orbits with inversion. In such a case, the explicitly
time-dependent observables 
$W_j(\ve{x},t) = W_{j}(\ve{x}) (-1)^{\eta(\ve{x},t)}$ should be defined
with an additional sign-factor which counts the number $\eta$ of 
inversions of the vector field along the trajectory,
$
(-1)^{\eta(\ve{\phi}_t(\ve{x}),t)}
\ve{e}_j(\ve{\phi}_t(\ve{x}))\cdot {\mat M}_t(\ve{x})\ve{e}_j(\ve{x}) > 0.
$
\bibitem{firstmoment} It is equivalent to show that $\int\dd^N\ve{x} \rho_0(\ve{x}) \dot{\ve{x}}(t) = 0$,
for finite but arbitrary long $t$, which follows from integrating by parts the rightmost side of eq. 
(\ref{eq:dyn}).
\end{thebibliography}
\end{document}